\documentclass[pdftex,twocolumn,epjc3]{svjour3}          
\usepackage{enumerate}
\RequirePackage[T1]{fontenc}

\smartqed  

\RequirePackage{graphicx}
\RequirePackage{mathptmx}      
\RequirePackage{flushend}
\RequirePackage[numbers,sort&compress]{natbib}
\RequirePackage[colorlinks,citecolor=blue,urlcolor=blue,linkcolor=blue]{hyperref}
\usepackage{amsmath}
\usepackage[caption=false]{subfig}

\journalname{Eur. Phys. J. C}

\begin{document}

\title{Spin-3/2 Dark Matter in a simple $t$-channel model}

\author{Mohammed Omer Khojali\thanksref{e1,ad1}, Ashok Goyal\thanksref{e2,ad2}, Mukesh Kumar\thanksref{e3,ad1} 
	\and
	Alan S. Cornell\thanksref{e4,ad1} 
}

\thankstext{e1}{e-mail: khogali11@gmail.com}
\thankstext{e2}{e-mail: agoyal45@yahoo.com}
\thankstext{e3}{e-mail: mukesh.kumar@cern.ch}
\thankstext{e4}{e-mail: Alan.Cornell@wits.ac.za}

\institute{National Institute for Theoretical Physics; School of Physics,
           University of the Witwatersrand, Johannesburg, Wits 2050, South Africa.\label{ad1}
           \and
           Department of Physics \& Astrophysics, University of Delhi, Delhi, India.\label{ad2}
}

\date{Received: date / Accepted: date}

\maketitle

\begin{abstract}
We consider a spin-3/2 fermionic dark matter (DM) particle interacting with the 
Standard Model quarks through the exchange of a charged and coloured scalar or 
vector mediator in a simple $t$-channel model.
It is found that for the vector mediator case, almost the entire parameter space allowed 
by the observed relic density is already ruled out by the direct detection LUX data. No 
such bounds exist on the interaction mediated by scalar particles. Monojet + missing
energy searches at the Large Hadron Collider provide the most stringent bounds 
on the parameters of the model for this case. 
The collider bounds put a lower limit on the allowed DM masses.
\end{abstract}
\begin{figure*}[t]
\centering
\subfloat[]{\includegraphics[height=165pt]{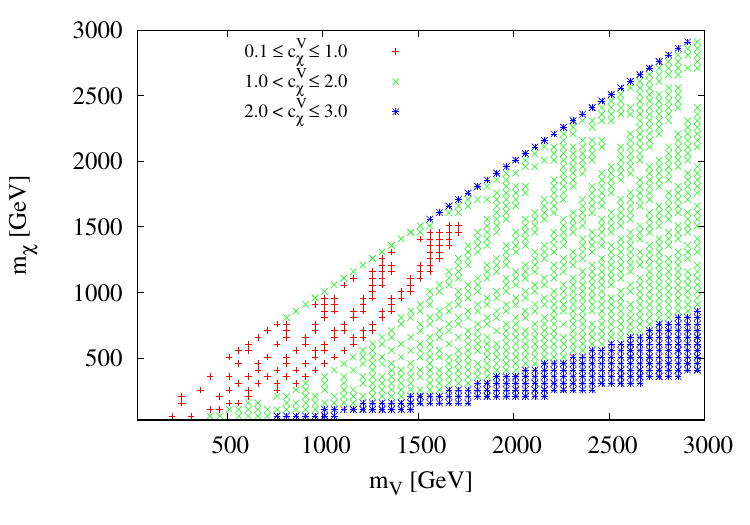}} \\
\subfloat[]{\includegraphics[height=165pt]{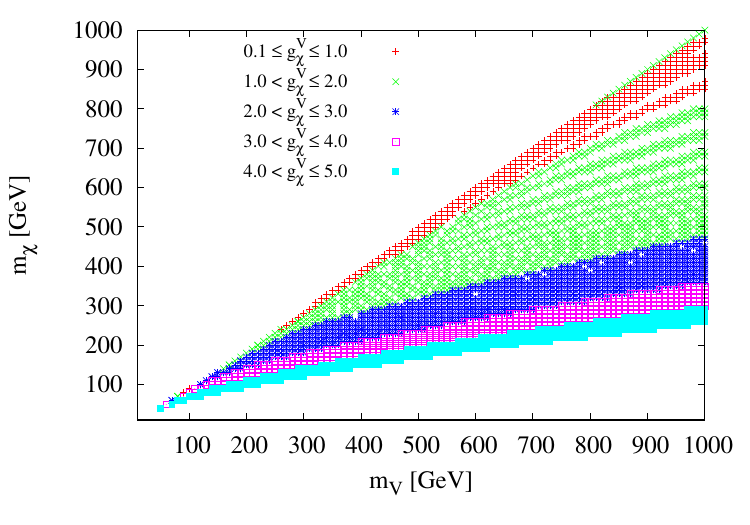}}
\subfloat[]{\includegraphics[height=165pt]{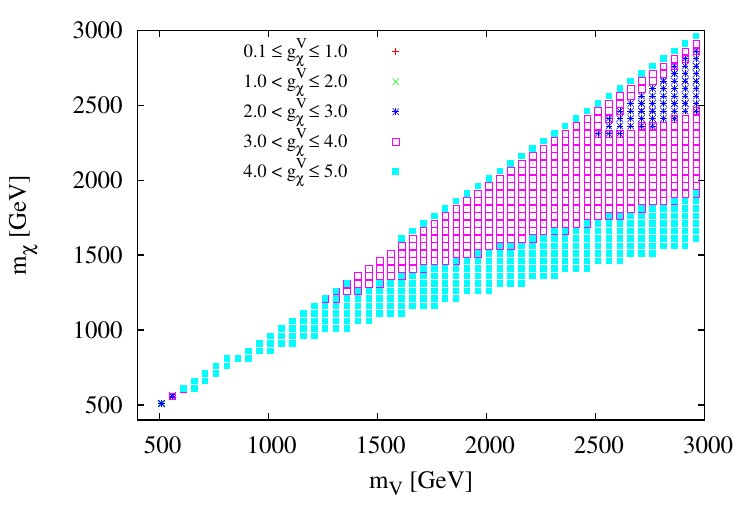}} \\
\subfloat[]{\includegraphics[height=165pt]{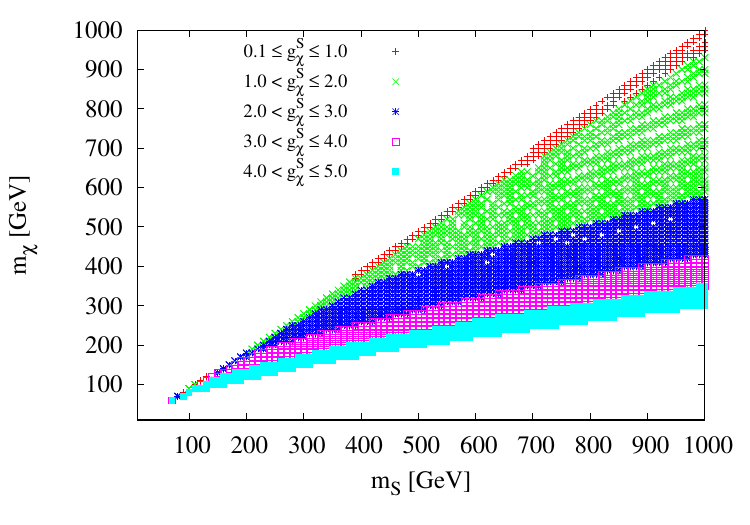}}
\subfloat[]{\includegraphics[height=165pt]{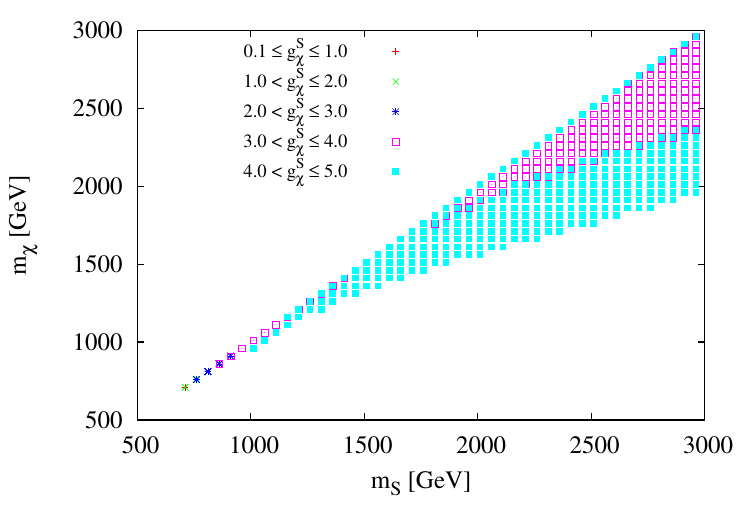}}
\caption{\label{Figure:1} Contour plots in the allowed DM mass $m_\chi$ and the mediator mass $m_V$/$m_S$. 
The color gradient corresponds to different values of the coupling. We have assumed that the DM $\chi$ saturates 
the observed relic density. The top panel corresponds to the dimension-4 interaction term for the vector mediator 
case. The middle and the bottom panels are for dimension-5 vector and scalar mediator cases respectively. The 
left panels correspond to a cut-off scale of 1~TeV and the right ones to the cut-off scale of 5~TeV for Yukawa couplings 
required to give the observed relic density. In the case of dimension-4 vector interaction, there is no cut-off required.}
\end{figure*}
\begin{figure*}[t]
\centering
\subfloat[]{\includegraphics[height=165pt]{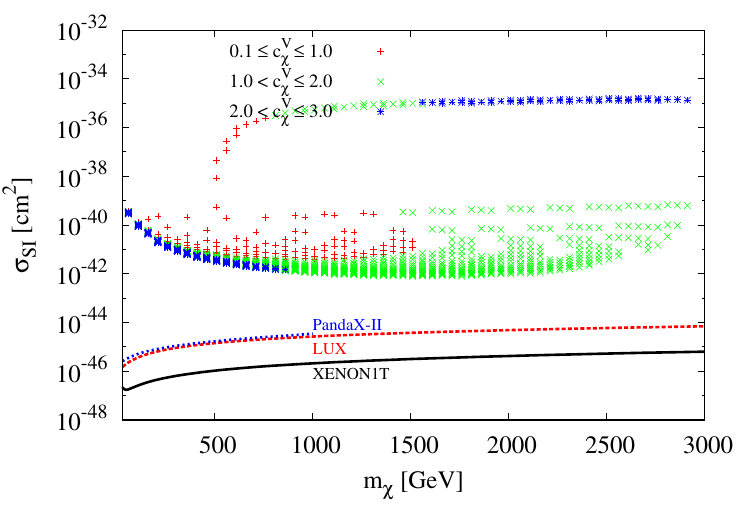}}\\
\subfloat[]{\includegraphics[height=165pt]{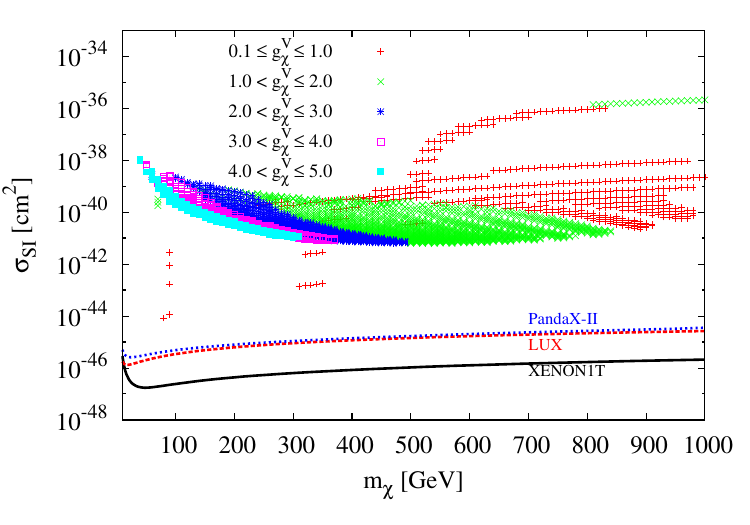}} 
\subfloat[]{\includegraphics[height=165pt]{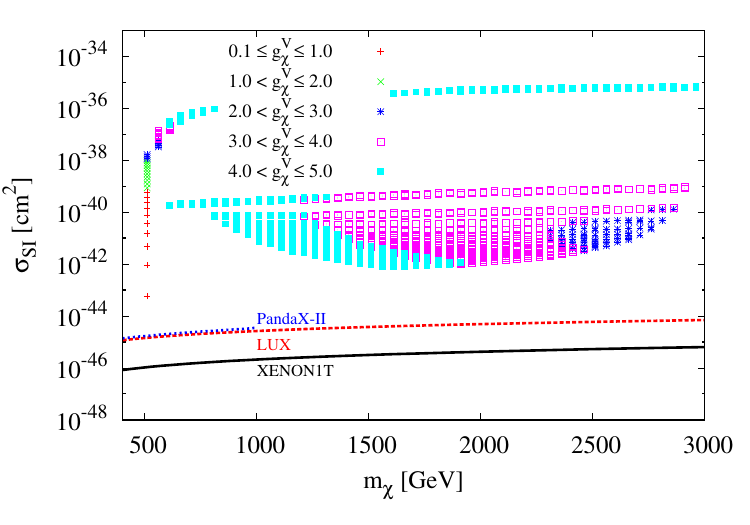}} 
\caption{\label{Figure:2} The spin-independent proton-DM cross-section $\sigma^{\rm SI}$. The top and the bottom panels correspond
to dimension-4 and dimension-5 vector interactions. The left and the right bottom panels correspond to the cut-off scales 1~TeV and 5~TeV 
respectively. The colour gradient stands for different Yukawa couplings. All parameters are consistent with the observed relic density. 
We have also shown the graphs from the observed current upper limits from LUX~\cite{Akerib:2016vxi} and PANDAX-II~\cite{Tan:2016zwf} 
experiments. The projected upper limit for XENON1T~\cite{Aprile:2015uzo} has also been shown. Almost the entire parameter space 
$(m_{\chi}, m_{V})$ for the vector mediator case considered here is already ruled out from the LUX data.}
\end{figure*}
\begin{figure*}[t]
\centering
\subfloat[]{\includegraphics[height=165pt]{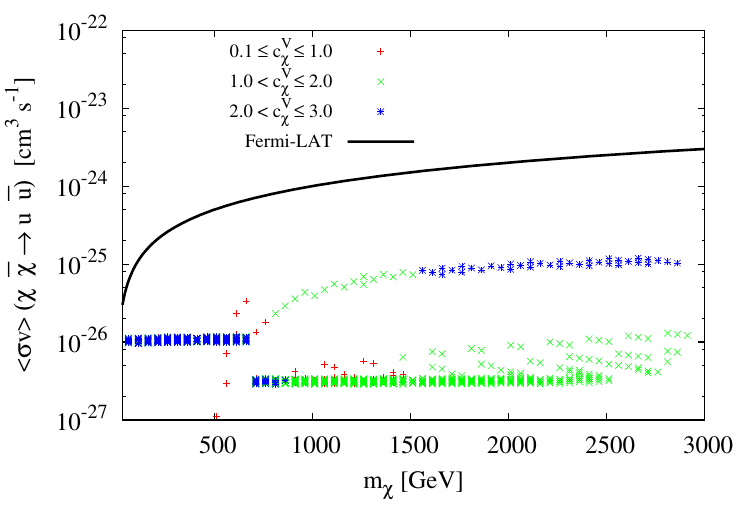}}\\
\subfloat[]{\includegraphics[height=165pt]{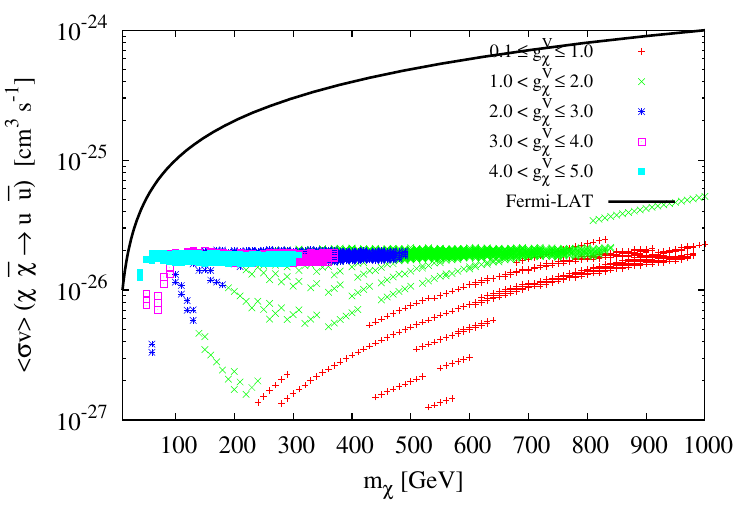}}
\subfloat[]{\includegraphics[height=165pt]{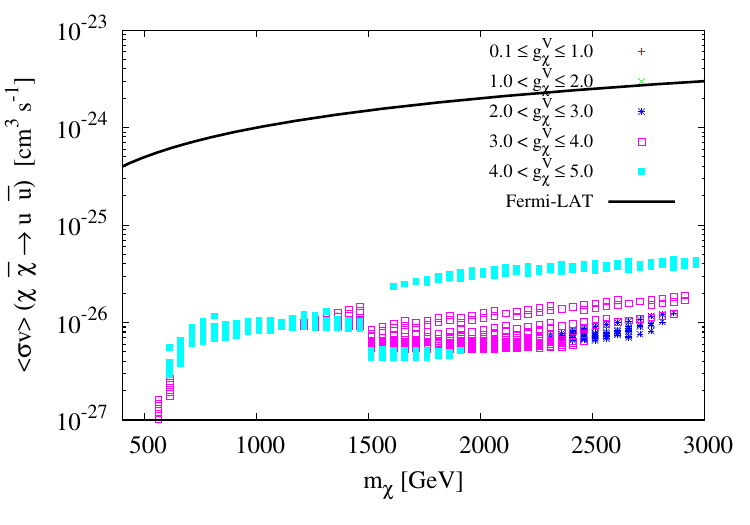}}\\
\subfloat[]{\includegraphics[height=165pt]{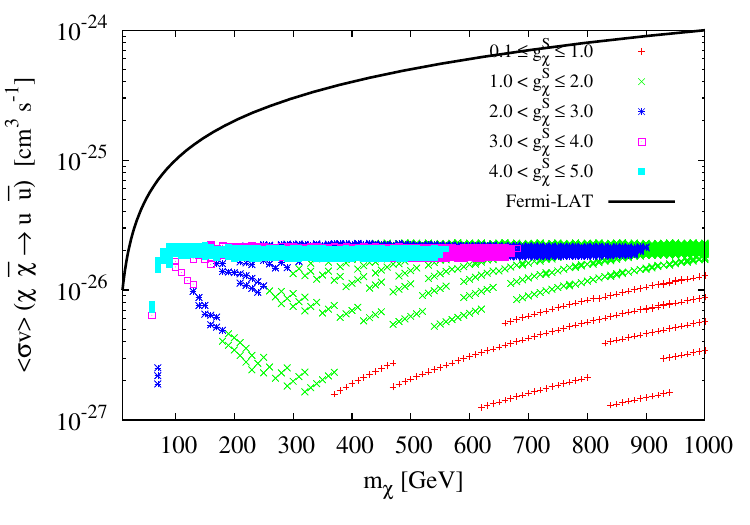}}
\subfloat[]{\includegraphics[height=165pt]{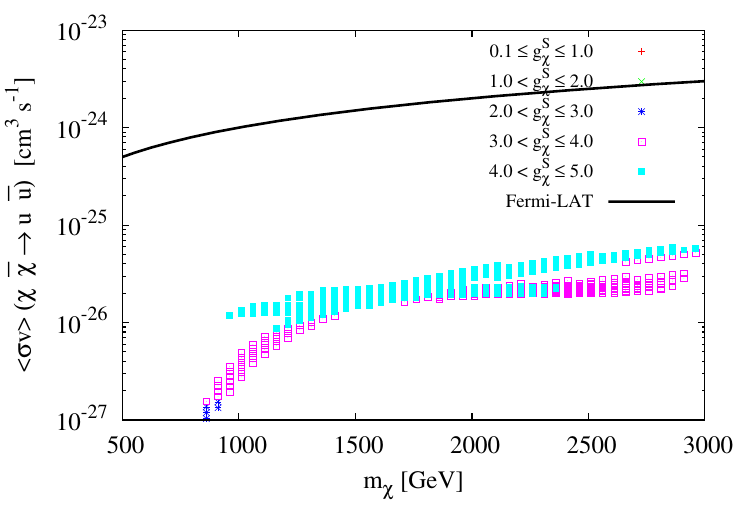}}
\caption{\label{Figure:3} The prediction for the DM $\chi\,\bar{\chi}$ annihilation rate into $u\,\bar{u}$, as a function of the DM mass $m_{\chi}$. 
All the parameters are chosen to be consistent with the observed relic density. The top  panel is for dimension-4 vector interaction. The middle 
and the bottom panels are for dimension-5 vector and  scalar interactions respectively. The left and the right panels  correspond to the cut-off 
scales 1~TeV and 5~TeV respectively. The colour gradient is for different values of the coupling as in other figures. Bounds from the Fermi-LAT 
experiments are also shown~\cite{Ackermann:2015zua}.}
\end{figure*}
\begin{figure*}[t]
\centering
\subfloat[]{\includegraphics[height=150pt]{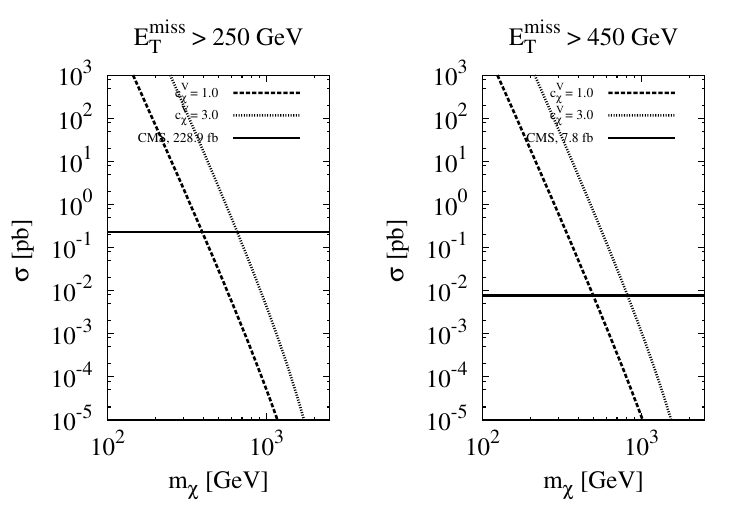}}\\
\subfloat[]{\includegraphics[height=150pt]{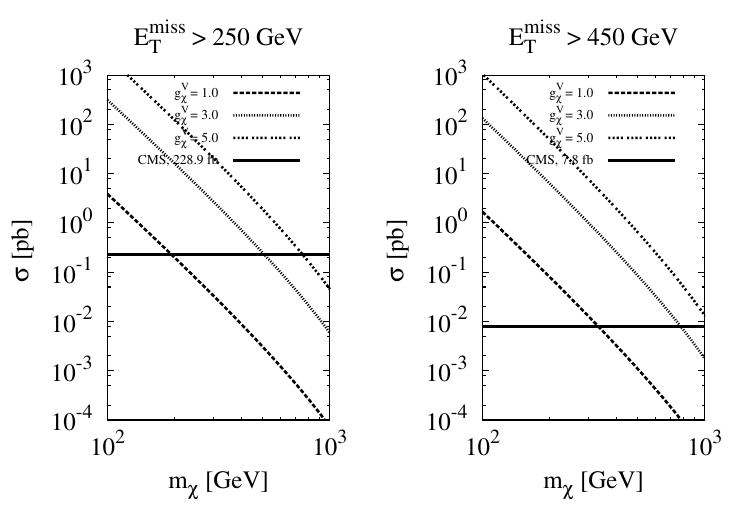}}
\subfloat[]{\includegraphics[height=150pt]{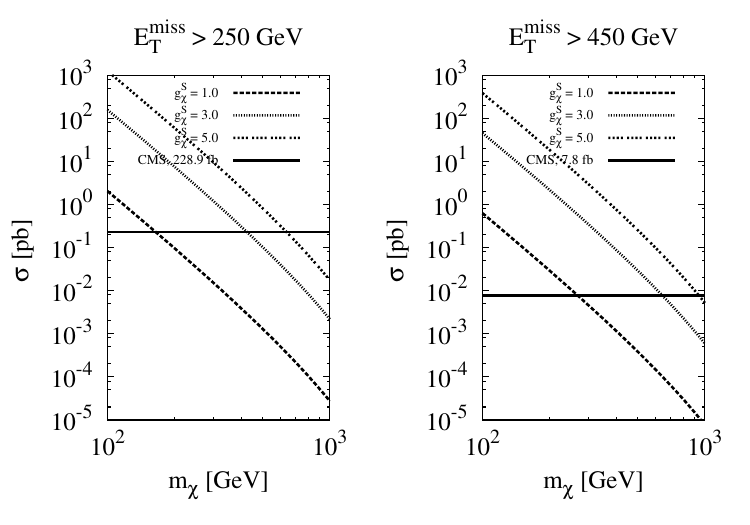}}
\caption{\label{Figure:4}The monojet cross-section in [pb] at the LHC with missing energy for two case viz. (i)  $E^{miss}_{T}>\,250\,$GeV 
and (ii) $E^{miss}_{T}>\,450\,$GeV. The cross-sections are obtained for all masses and couplings consistent with the observed relic density. 
Figures (a) and (b) corresponds to the dimension-4 interaction whereas Figures (b) and (c) correspond and dimension-5 vector and scalar 
interactions respectively at 1~TeV cut-off. The monojet cross-section from 8 TeV CMS collaboration data~\cite{Khachatryan:2014rra} based 
on an integrated luminosity 19.7\,fb$^{-1}$ is shown by solid black line.}
\end{figure*}

\section{Introduction}
\label{sec:intro}

Many astrophysical and cosmological observations during the last several decades provide strong evidence 
for the existence of dark matter (DM) in the Universe. The amount of DM has been precisely measured by the 
Planck satellite mission to be $\Omega_{\rm DM}h^{2}\,=\,0.1188\pm\,0.0010$~\cite{Ade:2015xua}, where 
the cold dark matter (CDM) content is estimated to comprise roughly $26\%$ of the total energy in the Universe. 
Investigations into the nature of dark matter particles and their interactions has emerged as an important field 
of research. Weakly interacting massive particle (WIMP) DM searches constitute an important programme 
at the Large Hadron Collider (LHC), where the ATLAS and CMS collaborations~\cite{Askew:2014kqa} are looking 
for DM signatures involving missing energy accompanied by a single or two jet events. It is expected, 
and there is indeed a real possibility, that the production of DM particles of any spin at 13\,TeV centre-of-mass energy 
would be detected.

Null results from the direct detection experiments~\cite{Akerib:2015cja,Aprile:2015uzo,Tan:2016zwf,Akerib:2016vxi}, 
which measure nuclear-recoil in DM-nucleon elastic scattering, have provided the most stringent upper bounds on 
the spin-independent DM-nucleon elastic scattering cross-section over a wide range of DM masses. This has 
provided important constraints on the DM models considered in the literature. In addition there are indirect detection 
experiments whose aim is to detect the signature of annihilating or decaying DM particles into the Standard Model (SM) 
particles.

Here we consider a spin-3/2 DM particle as an alternative to the conventional scalar, vector or spin-1/2 CDM particles. 
spin-3/2 CDM has been studied in EFT models and constraints from the relic density, direct and indirect observations 
obtained~\cite{Yu:2011by,Ding:2012sm,Ding:2013nvx,Savvidy:2012qa}. spin-3/2, 7.1\,KeV warm dark matter (WDM) 
has been considered as a means to provide 
a viable explanation from the anomalous 3.1\,KeV X-ray line observed by the XMM Newton~\cite{Dutta:2015ega}. 
Furthermore spin-3/2 DM with a Higgs portal has also been recently investigated~\cite{Chang:2017dvm}. This idea of using
a spin-3/2 DM candidate was also considered in our recent paper~\cite{Khojali:2016pvu}, where we considered a minimal 
SM singlet spin-3/2 DM candidate interacting with a spin-1 mediator in a minimal flavour violation (MFV) $s$-channel model. 
In the case of pure vector interactions, it was found that almost the entire parameter space was allowed by the observed relic 
density, and is ruled out by the direct detection observation data.

In this paper we consider a minimal spin-3/2 SM singlet DM candidate. The DM particle in this case is a $t$-channel 
annihilator. It interacts with the SM particles through the exchange of a scalar (S) or vector (V) particle. A class of 
such models for scalar and vector mediator couplings with a spin-1/2 DM candidate has been considered in 
Refs.~\cite{Chang:2013oia, DiFranzo:2013vra, An:2013xka, Papucci:2014iwa}. 
The mediators in these $t$-channel models carry colour 
or leptonic index. As such we shall describe the model for this study in section~\ref{sec:Model}, and in 
section~\ref{sec:constraints} we discuss all relevant experimental constraints. The relic density contributed by the 
DM particles is calculated (taking into account the co-annihilation processes), 
and assuming that the contribution by these spin-3/2 DM particles does not exceed the 
observed relic density, constraints on the parameters of the model are obtained in section~\ref{sec:relic}. With these 
constraints  in place we discuss the compatibility of these constraints from the direct and indirect detection experiments 
in section~\ref{sec:direct} and section~\ref{sec:indirect} respectively. In section~\ref{sec:collider} we examine the 
signature of these DM particles at the LHC, where a monojet signal with missing energy is investigated. 
Section~\ref{sec:summary} is devoted to the summary and discussion of our main results.              

\section{The Model}
\label{sec:Model}
The model consists of a SM singlet spin-3/2 particle interacting through the mediation of a scalar (S) or a 
vector ($V^{\mu}$) which carries a baryonic (colour) or lepton index. In general the mediator couples to 
right-handed up-type quarks (or leptons), right-handed down-type quarks (or leptons) or left-handed quark 
(or lepton) doublets. We consider here the right-handed up-type quark case for simplicity, where the other 
cases are similar. The spin-3/2 free Lagrangian is given by~\cite{Christensen:2013aua}:
\begin{equation}
\mathcal{L} = \bar{\chi}_{\mu} \Lambda^{\mu\nu} \chi_{\nu},
\end{equation}
with
\begin{equation}
\Lambda^{\mu\nu} = (i{\not}\partial-m_{\chi})g^{\mu\nu} - i(\gamma^{\mu}\partial^{\nu} + \gamma^{\nu} \partial^{\mu})
+ i\gamma^{\mu}{\not}\partial\gamma^{\nu} + m_{\chi}\gamma^{\mu}\gamma^{\nu}.
\end{equation}
Note that \(\chi_{\mu}\) satisfies \(\Lambda^{\mu\nu}\chi_{\nu} = 0\), and with $\chi_\mu$ being on mass-shell we have 
\begin{equation}
(i{\not}\partial-m_{\chi})\chi_{\mu} = \partial^{\mu} \chi_{\mu} = \gamma^{\mu}\chi_{\mu} = 0.
\end{equation} 
The spin sum for spin-3/2 particles are 
\begin{equation}
S_{\mu\nu}^{+}(p) = \sum_{i = -3/2}^{3/2} u^{i}_{\mu}(p)\bar{u}^{i}_{\nu}(p),
\end{equation}
\begin{equation}
S_{\mu\nu}^{-}(p) = \sum_{i = -3/2}^{3/2} v^{i}_{\mu}(p)\bar{v}^{i}_{\nu}(p),
\end{equation}
and are given by~\cite{Christensen:2013aua}:
\begin{align}
S^{\pm}_{\mu\nu}(p) =& -({\not}p \pm m_{\chi}) \Big[g_{\mu\nu} - \frac{1}{3}\gamma_{\mu}\gamma_{\nu} 
-\frac{2}{3m^{2}_{\chi}}{\not} p_{\mu}{\not} p_{\nu}
\nonumber\\
&\mp\frac{1}{3m_{\chi}}(\gamma_{\mu}p_{\nu} - \gamma_{\nu}p_{\mu})\Big].
\end{align}

In view of the non-renormalisable nature of interacting spin-3/2 theories, we can only write generic interactions 
which respect to the SM gauge symmetry between the singlet, $\chi$, with SM fermions mediated by a scalar 
or a vector~\cite{Hassanain:2009at}. The effective theory is endowed with a cut-off $\Lambda$ and breaks down at high energies. 
However, it is safe to consider the effective theory as long as the momentum transfer stays below the cut-off and the DM and 
mediator masses remains less than the cut-off.
In order to illustrate the effect of the cut-off scale we have chosen two values of the cut-off, scale namely $\Lambda = 1$~TeV and 
5~TeV. The dimension-4 vector interaction is independent of the cut-off scale $\Lambda$.  Its coupling however, is 
restricted by the perturbative limit. We will consider the vector and scalar mediator cases separately:
\begin{enumerate}[{(a)}]
\item Scalar mediator $S$:
For the scalar mediator case we can write the SM gauge invariant interaction as: 
\begin{equation}\label{eq:S 4D int}
\mathcal{L}_{int}\,\supset\,-\frac{\left(g_{\chi}^{\rm S}\right)^{i}}{\Lambda}\,\bar{\chi}_{\mu}\,g^{\mu\nu}\,u^{i}_{R}\,D_{\nu} S^{*}_{i} + \rm{h.c.},
\end{equation}
where $i$ is a generation index and $u^{i}_{R}\,\equiv\, (u_{R},\,c_{R},\,t_{R})$.  In this case we do not have a 
dimension-4 interaction term. This is because of the nature of the vector-spinor $\chi_{\mu}$, which on the 
mass-shell satisfies $\gamma^{\mu}\,\chi_{\mu}\,=\,0$, and thus it is not possible to construct a Lorentz-invariant 
dimension-4 interaction term involving $\chi_{\mu}$,\, $S$ and the Dirac field $u_{R}$.
\end{enumerate}

\begin{enumerate}[{(b)}]
\item Vector mediator $V_{\mu}$:
In this case we can write dimension-4 as well as dimension-5 interaction terms, namely
\begin{equation}\label{eq:V 4D int}
\mathcal{L}_{int}\,\supset\, i\,\left(c_{\chi}^{\rm V}\right)^{i}\,\bar{\chi}_{\mu}\,u^{i}_{R}\,(V^{\mu}_{i})^{*} + \rm{h.c.}
\end{equation}
and 
\begin{equation}\label{eq:V 5D int}
\mathcal{L}_{int}\,\supset\, i\frac{\left(g_{\chi}^{\rm V}\right)^{i}}{\Lambda}\,\bar{\chi}_{\mu}\,g^{\mu\alpha}\,\gamma^{\beta}\,u^{i}_{R}\,V^{*i}_{\alpha\beta}\,+ \rm{h.c.}\,.
\end{equation}
\end{enumerate}
For all calculations we set $\Lambda = 1$~TeV. 
The interaction Lagrangian for the scalar and vector can be written as:
\begin{equation}
\mathcal{L}_{scalar}\,=\,(D_{\mu}\, S_{i})^{\dagger}(D^{\mu}\,S_{i})\,-\,m^{2}_{S_{i}}\,S^{\dagger}_{i}\,S_{i},
\end{equation}
\begin{equation}
\mathcal{L}_{vector}\,=\,-\frac{1}{4}V^{\dagger i}_{\mu\nu}\,V_{i}^{\mu\nu}\,+\,m^{2}_{V}\,V^{\dagger}_{\mu\,i}\,V^{\mu\,i}\,+\,i\,g_{s}\,V^{\dagger}_{\mu\,i}\,t^{a}\,V^{i}_{\mu}\,G^{\mu\nu}_{a},
\end{equation}
where $V^{i}_{\mu\nu}\,=\,D_{\mu}\,V^{i}_{\nu}\,-\,D_{\nu}\,V^{i}_{\mu}$.
The covariant derivative is given by 
\begin{equation}
D_{\mu}\,=\,\partial_{\mu}\,+\,i\,g_{s}\,t_{a}\,G^{a}_{\mu}\,+i\,g\,\frac{1}{2}\,\vec{\tau}\,.\,\vec{W_{\mu}}\,+i\,g^{\prime}\,\frac{1}{2}\,Y\,B_{\mu},
\end{equation}
where $g_{s}$ is the QCD strong coupling constant. Unlike the $s$-channel mediator, where a single mediator is required, 
in the $t$-channel model we require a different mediator for each generation.  

In general, the interaction given in Lagrangian~(\ref{eq:S 4D int}),~(\ref{eq:V 4D int}) and~(\ref{eq:V 5D int})
induce flavour-changing neutral currents (FCNC), which are strongly constrained by low energy phenomenology. 
The FCNC constraints can be avoided by imposing a minimal flavour violation (MFV) structure on the Yukawa couplings. 
The parameter space will consist of the DM candidates mass $m_\chi$, the vector (scalar) couplings 
 $\left(c_{\chi}^{\rm V}\right)^{i},\,\left(g_{\chi}^{\rm V}\right)^{i}$, $\left(\left(g_{\chi}^{\rm S}\right)^{i}\right)$ and the mediator 
masses $m^{i}_{V}$ ($m^{i}_{S}$), for each generation. 
For simplicity we will set the couplings and mediator masses for all the generations to be equal. If the mediator mass in 
the kinematically accessible region of the LHC, the decay of the mediator and the ensuing signal will become important. 
The decay width of the scalar and vector mediators
$\Gamma(S^{i}\,/V^{i}\,\rightarrow\,\chi\,\bar{u}_{i})$, dropping the generation index, are given by:
\begin{align}
\Gamma(S\,\rightarrow\,\chi\,\bar{u}) =& \frac{\left(g^{\rm S}_{\chi}\right)^{2}\,m^{5}_{S}}{96\,\pi\,\Lambda^{2}\,m^{2}_{\chi}}\Bigg[1 -\Bigg(\frac{m_{\chi}}{m_{S}} + \frac{m_{u}}{m_{S}}\bigg)^{2}\Bigg]\,
\nonumber\\&\times\Bigg[1 -\Bigg(\frac{m_{\chi}}{m_{S}} - \frac{m_{u}}{m_{S}}\bigg)^{2}\Bigg]
\times\Bigg[1\,-\,\frac{m^{2}_{\chi}}{m^{2}_{S}}\,-\,\frac{m^{2}_{u}}{m^{2}_{S}}\Bigg]\nonumber\\&
\times \,\lambda^{1/2}\Bigg(1,\,\frac{m^{2}_{\chi}}{m^{2}_{S}},\,\frac{m^{2}_{u}}{m^{2}_{S}}\Bigg)\nonumber\\&\simeq\,\frac{\left(g^{\rm S}_{\chi}\right)^{2}\,m^{5}_{S}}{96\,\pi\,\Lambda^{2}\,m^{2}_{\chi}} \Bigg(1 - \frac{m^{2}_{\chi}}{m^{2}_{S}}\Bigg)^{4}, 
\end{align}
since $m^{i}_{S}$,\,$m_{\chi}$\,$\gg$\,$m_{u}$ is true for all quarks, except the top quark, and $\lambda(a,\,b,\,c)\,\equiv a^{2}\,+\,b^{2}\,+\,c^{2}\,-\,2\,a\,b\,-\,2\,a\,c\,-\,2b\,c$\,; 
\begin{align}
\Gamma(V\,\rightarrow\,\chi\,\bar{u}) =& \frac{\left(c^{\rm V}_{\chi}\right)^{2}\,m_{V}}{288\,\pi}\Bigg(1\,-\,\frac{m^{2}_{\chi}}{m^{2}_{V}}\,-\,\frac{m^{2}_{u}}{m^{2}_{V}}\Bigg)\Bigg[5\,+\,\frac{m^{2}_{V}}{m^{2}_{\chi}} \nonumber\\
&+\, \frac{m^{2}_{\chi}}{4\,m^{2}_{V}}\, -\,\frac{m^{2}_{u}}{m^{2}_{\chi}}-\,\frac{m^{2}_{u}}{2\,m^{2}_{V}}
\,+\, \frac{m^{4}_{u}}{4\,m^{2}_{V}\,m^{2}_{\chi}}\Bigg] \nonumber\\
&\qquad \times\lambda^{1/2}\Bigg(1,\,\frac{m^{2}_{\chi}}{m^{2}_{V}},\,\frac{m^{2}_{u}}{m^{2}_{V}}\Bigg)\nonumber\\
&\simeq \frac{\left(c^{\rm V}_{\chi}\right)^{2}\,m_{V}}{288\,\pi}\Bigg(1\,-\,\frac{m^{2}_{\chi}}{m^{2}_{V}}\Bigg)^{2}  \nonumber\\
&\qquad \times \Bigg(5\,+\,\frac{m^{2}_{V}}{m^{2}_{\chi}}\,+\, \frac{m^{2}_{\chi}}{4\,m^{2}_{V}}\Bigg),
\end{align} 
and
\begin{align}
\Gamma(V\,\rightarrow\,\chi\,\bar{u})=& \frac{\left(g^{\rm V}_{\chi}\right)^{2}\,m^{5}_{V}}{288\,\pi\,\Lambda^{2}\,m^{2}_{\chi}}\Bigg[\frac{m^{2}_{\chi}}{m^{2}_{V}}\,+\, \frac{m^{4}_{\chi}}{m^{4}_{V}}\,-\,\frac{3\,m^{6}_{\chi}}{m^{6}_{V}}\,\nonumber\\&+\,\Bigg(1\,-\,\frac{m^{2}_{u}}{m^{2}_{V}}\Bigg)^{3}
+\,\frac{5\,m^{4}_{\chi}\,m^{2}_{u}}{m^{6}_{V}}\,-\,\frac{m^{2}_{\chi}\,m^{4}_{u}}{m^{6}_{V}}\Bigg]\nonumber\\&\times\lambda^{1/2}\Bigg(1,\,\frac{m^{2}_{\chi}}{m^{2}_{V}},\,\frac{m^{2}_{u}}{m^{2}_{V}}\Bigg)
\simeq\,\frac{\left(g^{\rm V}_{\chi}\right)^{2}\,m^{5}_{V}}{288\,\pi\,\Lambda^{2}\,m^{2}_{\chi}}\nonumber\\&\times\Bigg[1\,+\,\frac{m^{2}_{\chi}}{m^{2}_{V}}\,+\,\frac{m^{4}_{\chi}}{m^{4}_{V}}\,-\,\frac{3\,m^{6}_{\chi}}{m^{6}_{V}}\Bigg]\,\Bigg(1\,-\,\frac{m^{2}_{\chi}}{m^{2}_{V}}\Bigg), 
\end{align} 
for dimension-4 and dimension-5 interaction Lagrangians given in Eqs.~(\ref{eq:V 4D int}) and~(\ref{eq:V 5D int}) respectively.

We notice that the decay width of the scalar/vector mediators become large as $m_{\chi}$ decreases. This happens because of the existence of inverse power of $m_{\chi}^{2}$ terms in spin-3/2 particle polarization sum. Since $\Gamma\left(\rm S/V\right)\,<\,m_{\rm S/V} $ is required for the mediator description to be perturbatively valid $m_{\chi}^{2}\,>\, \left(g_{\chi}^{\rm S} \right)^{2}m_{\rm S}^{4} / \left(96\,\pi\,\Lambda^{2}\right)$ and $m_{\chi}^{2}\,>\, \left(g_{\chi}^{\rm V} \right)^{2} m_{\rm V}^{4}/\left(288\,\pi\,\Lambda^{2}\right)$ for dimension-5 scalar and vector interactions respectively.

\section{Constraints}
\label{sec:constraints}
In this section we examine the constraints on the model parameters $m_{\chi}$,\,$m_{S}$,\,$m_{V}$ and the coupling 
constants from the relic density, direct and indirect observations. 
\subsection{Relic density}
\label{sec:relic}
  In the early Universe the DM relic density is determined by the dominant DM annihilation  processes
 $\chi\,\bar{\chi}\,\rightarrow\,u\,\bar{u}$, mediated by the $t$-channel exchange of scalar/vector mediators. 
 Since the mediators in this model carry colour and charge, co-annihilation processes like $\chi\,S(V)\,\rightarrow\,u\,g$ and 
 $S\,S^{*}\,(V\,V^{*})\,\rightarrow\,g\,g$ (even though exponentially suppressed when mass splitting 
 $\left(m_{S/V}\,-\,m_{\chi}\right) >$\, freeze-out temperature $T_{f}$), will play an important role if the DM mass gets close 
 to the mediator mass. The co-annihilation processes $\chi\,S(V)\,\rightarrow\,u\,g$ are mediated by $t$-channel exchanges of mediators,
 as well as by $s$-channel exchanges of gluons and through the four-point interaction involving the DM, mediator, $u$-quark and the 
 gluon vertex. These processes will reduce the Yukawa coupling needed to generate the required thermal relic abundance. Self 
 annihilation mediator processes $S\,S^{*}(V\,V^{*})\,\rightarrow\,g\,g$ are generated by purely gauge interactions, and are 
 independent of the Yukawa couplings, having the potential to suppress the relic density below the observed value.
 
 At freeze-out the DM and mediator particles are non-relativistic. In the non-degenerate parameter space the channel 
 $\chi\,\bar{\chi}\,\xrightarrow{S/V}\,u\,\bar{u}$ cross-section can be easily evaluated, and in the limit 
 $m_{\chi},\,m_{S},\,m_{V}\,\gg\,m_{u}$ are given by 
 \begin{align}
 \langle\sigma(\chi\,\bar{\chi}\,\xrightarrow{S}\,u\,\bar{u})|v\rangle\,\simeq\,
 \frac{\left(g_{\chi}^{\rm S}\right)^{4}\,m_{\chi}^{2}}{768\,\pi\,\Lambda^{4}}\frac{1}{\left(1\,+\,r^{2}\right)^{2}},
\end{align}
\begin{align}
 \langle\sigma(\chi\,\bar{\chi}\,\xrightarrow{V}\,u\,\bar{u})|v\rangle\,\simeq&\, \frac{\left(c_{\chi}^{\rm V}\right)^{4}}{1536\,\pi\,m_{\chi}^{2}}
 \frac{1}{\left(1\,+\,r^{2}\right)^{2}}  \nonumber\\
 &\quad \times \left[5\,-\,\frac{4}{r^{2}}\,+\,\frac{2}{r^{4}}\right],
 \end{align}
and 
\begin{align}
 \langle\sigma(\chi\,\bar{\chi}\,\xrightarrow{V}\,u\,\bar{u})|v\rangle\,\simeq&\,
 \frac{\left(g_{\chi}^{\rm V}\right)^{4}\,m_{\chi}^{2}}{768\,\pi\,\Lambda^{4}}\frac{1}{\left(1\,+\,r^{2}\right)^{2}}  \nonumber\\
 &\quad \times \left[5\,+\,\frac{1}{24}\frac{1}{\left(1\,+\,r^{2}\right)^{2}}\right],
\end{align}
for the scalar-mediator and the vector-mediator dimension-4 and dimension-5 interaction 
Lagrangians~(\ref{eq:S 4D int}),~(\ref{eq:V 4D int}) and~(\ref{eq:V 5D int}) respectively, with the mass ratio 
$r\,=\,m_{S}(m_V) / m_{\chi}$. 
 The thermal relic density of $\chi$'s is obtained by solving the Boltzmann equation:    
\begin{align}
\frac{d \eta_{\chi}}{d t} + 3 H \eta_{\chi} = -\langle\sigma |v\rangle \left(\eta^{2}_{\chi} - (\eta^{\rm eq}_{\chi})^{2}\right),
\end{align}
where $H$ is the Hubble constant, $\langle\sigma |v\rangle$ is the thermally averaged $\chi$-annihilation cross-section, 
and 
\begin{align}
\eta^{eq}_{\chi} = 4\left(\frac{m_{\chi}\, T}{2\pi}\right)^{3/2} \exp\left(-\frac{m_{\chi}}{T}\right).
\end{align}
The freeze out occurs when the $\chi$'s are non-relativistic, and then $\langle\sigma |v\rangle$ can be written as: 
\begin{align}
\sigma|v|\,=\,a\,+\,b\,v^{2}\,+\,\mathcal{O}(v^{4}). 
\end{align}  
The Boltzmann equation can be solved numerically to yield following~\cite{E. W. Kolb}:
\begin{align}
\Omega_{\rm DM} h^{2}\simeq \frac{2\times1.07\times10^{9}X_{F}}{M_{pl}\sqrt{g_{*}}(a + \frac{3b}{X_{F}})},
\end{align}
where \(g_{*}\) is the number of degrees of freedom at freeze-out temperature \(T_{F}\), \(X_{F} = m_{\chi}/{T_{F}}\) is 
obtained by solving
\begin{align}
X_{F} = \ln\Bigg[c(c+2)\sqrt{\frac{45}{8}}\frac{g M_{pl} m_{\chi}(a + \frac{6b}{X_{F}})}{2\pi^{3}\sqrt{g_{*}(X_{F})}\sqrt{X_{F}}}\Bigg],
\end{align}
where $c$ is taken to be 1/2. $g_{*}(X_{F})$ is the number of degrees of freedom at the freeze-out 
temperature, and is taken to be 92 for our estimate, and $g\,=\,4$.

The annihilation cross-section for the co-annihilation processes $\chi\,S(V)\,\rightarrow\,u\,g$ in this limit are given by
\begin{equation}
 \langle\sigma(\chi\,S\,\rightarrow\,u\,g)|v\rangle\,\simeq\,\frac{\left(g_{\chi}^{\rm S}\right)^{2}g_{s}^{2}}{64\,\pi\,\Lambda^{2}}\,
 \frac{(1\,+\,r)}{r^{3}}\left[1\,+\,\frac{14}{9}\,r\,+\,\frac{13}{27}\,r^{2}\right], 
\end{equation}
\begin{align}
  \langle\sigma(\chi\,V\,\rightarrow\,u\,g)|v\rangle\,\simeq&\frac{\left(c_{\chi}^{\rm V}\right)^{2}\,g_{s}^{2}}{165888\,\pi\,m_{\chi}^{2}}\,
  \frac{1}{r^{6}(1\,+\,r)}\Bigg[1164\,+\,5628\,r\nonumber\\&+\, 11403\,r^{2}+\, 12568\,r^{3}\,+\,8242\,r^{4}\nonumber\\&
  +\,2452\,r^{5}\,+\,
  319\,r^{6}\Bigg],
\end{align}
and 
\begin{align}
  \langle\sigma(\chi\,V\,\rightarrow\,u\,g)|v\rangle\,\simeq&\frac{\left(g_{\chi}^{\rm V}\right)^{2}\,g_{s}^{2}}{497664\,\pi\,\Lambda^{2}}\,
  \frac{1}{r^{5}(1\,+\,r)}\Bigg[372\,+\,2724\,r\nonumber\\&+\, 6537\,r^{2}+\,8742\,r^{3}\,+\,7072\,r^{4}\,+\nonumber\\&5222\,r^{5}\,+\,307\,r^{6}\Bigg].
\end{align}

To calculate the relic density we have implemented the $t$-channel scalar and vector interactions with SM 
quarks and spin-3/2 DM, including the relevant co-annihilation processes, in \texttt{micrOMEGAS}~\cite{Belanger:2014vza}.
Note that this numerically solves the Boltzmann equation by taking the full expressions of the annihilation cross-section. 
We have checked the relic abundance in the non-degenerate parameter space for some representative values of the parameters, 
and found them to be in agreement with the numerical calculations done by the package \texttt{micrOMEGAS}. The necessary model files 
for \texttt{micrOMEGAS} were built using \texttt{FeynRules}~\cite{Alloul:2013bka}. 
In Figure~\ref{Figure:1} we show the contour graphs in the DM-mediator mass plane. The colour gradients correspond to the 
Yukawa couplings to conform to the observed relic density density $\Omega_{\rm DM}h^{2}\,\simeq\,0.12$. 
In the top panel we have shown the allowed values of the DM and mediator masses for the dimension-4 interaction term for the 
vector interaction. This case is not dependent on the cut-off scale and we find that co-annihilation is important for DM 
particle masses up to roughly 1.5~TeV. The middle and the bottom panels are for dimension-5 vector and scalar mediator cases 
respectively. The left panels correspond to a cut-off scale of 1~TeV and the right ones to the cut-off scale of 5~TeV. 
For 1~TeV cut-off scale, we observed that co-annihilation channel is becoming important as compared to the self-annihilation processes 
for all DM masses.
We have restricted the vector couplings within the perturbative limit namely $c_\chi^V < \sqrt{4 \pi}$. In the case of the 5~TeV cut-off scale, as can 
be seen from the right hand middle and bottom panels, co-annihilation does not seem to play any significant role. In the parameter space in which 
co-annihilation is not important, comparatively large Yukawa couplings are required to obtain the required relic density.

\subsection{Direct detection}
\label{sec:direct}
Direct detection experiments~\cite{Tan:2016zwf, Akerib:2016vxi, Akerib:2015cja, Aprile:2015uzo} on elastic 
nucleon-DM scattering have provided the most stringent bounds on DM mass and interactions in a large 
number of conventional DM models. In the $t$-channel spin-3/2 DM model considered here, the cross-sections 
at zero momentum transfer can be easily calculated~\cite{DelNobile:2013sia,Freytsis:2010ne, Barger:2008qd}. 
The dominant contribution to the spin independent DM-nucleon scattering cross-section is estimated by noting 
that there are two important scales that are present in the scattering. There is a cut-off scale $\Lambda$ at the 
TeV scale and the QCD scale\,$\sim$\,100 MeV, which represents the typical energy and momentum of quarks 
bound inside the non-relativistic nucleons. In the leading order, neglecting the quark momenta, the DM-nucleon 
scattering amplitude for the dimension-5 vector interaction, for example, proceeds through the $s$-channel exchange 
and is given by 
\begin{align}
M\,\simeq &\frac{1}{2}\,\frac{\left(g_{\chi}^{\rm V}\right)^{2}\,m_{\chi}^{2}}{\Lambda^{2}}\,\bar{u}^{\alpha}\left(p_{2}\right)\,
\gamma^{\mu}\,P_{R}\,u^{\beta}\left(p_{1}\right)\,\frac{\left(g_{\alpha\,\beta}\,-\,\frac{p_{\alpha}\,p_{\beta}}{m_{{\rm V}}^{2}} \right)}{s\,-\,m_{{\rm V}}^{2}}\nonumber\\&\hspace{0.5cm}\times\,\bar{v}\left(p_{3}\right)\,\gamma_{\mu}\,P_{R}\,v\left(p_{4}\right),
\end{align}
where $p_{1}$ and $p_{2}$ are the momenta of the incoming and outgoing DM particles, $p_{3}$ and $p_{4}$ are the corresponding incoming and outgoing quark momenta and $p\,=\,p_{1}\,+\,p_{3}$. 
In the non-relativistic approximation 
\begin{align}
 M\,\simeq&\,\frac{1}{8}\,\frac{\left(g_{\chi}^{\rm V}\right)^{2}\,m_{\chi}^{2}}{\Lambda^{2}}\,\frac{1}{m_{\chi}^{2}\,-\,m_{{\rm V}}^{2}} \nonumber \\
 &\quad \times \left[\bar{u}^{\alpha}\left(p_{2}\right)\gamma^{\mu}\,u_{\alpha}\left(p_{1}\right)\,\bar{v}\left(p_{3}\right)\,\gamma_{\mu}\,v\left(p_{4}\right)\right].
\end{align}
The DM-nucleon scattering cross-section can now be easily calculated and we get 
\begin{equation}~\label{eq:cross5D}
\sigma^{\rm SI}\,\simeq\, \frac{1}{64\,\pi}\left(\frac{g_{\chi}^{\rm V}}{\Lambda}\right)^{4}\frac{\mu^{2}}{\left(1\,-\,r^{2} \right)^{2}}\,f_{N},
\end{equation}
and similarly for the dimension-4 vector mediated interaction, we get 
\begin{equation}~\label{eq:cross4D}
\sigma^{\rm SI}\,\simeq\, \frac{1}{64\,\pi}\frac{\left(c_{\chi}^{\rm V}\right)^{2}\,\mu^{2}}{m_{\chi}^{4}\,\left(1\,-\,r^{2} \right)^{2}}\,f_{N},
\end{equation}      
where $\mu\,=\,m_{\chi}\,m_{N}/(m_{\chi}\,+\,m_{N})$, $f_{N}\,=\,4$ for protons and 1 for the neutrons, 
and we have dropped the terms proportional to the quark mass and momenta in comparison to the leading term.
The elastic nucleon-DM cross section for the case of scalar mediator is 
suppressed by terms proportional to quark momenta and has not been considered here.

In Figure~\ref{Figure:2} we show the predictions for the spin-independent DM-proton scattering 
cross-sections, $\sigma^{\rm SI}$, for the vector mediator case. 
The top and the bottom panels correspond to dimension-4 and dimension-5 vector interactions respectively. 
The left and the right bottom panels correspond to the cut-off scales 1~TeV and 5~TeV respectively. 
The colour gradient stand for different Yukawa couplings where all parameters are consistent with the 
observed relic density. We have also shown the observed current upper limits from 
LUX~\cite{Akerib:2016vxi} and PANDAXII~\cite{Tan:2016zwf} experiments, as well as the projected upper limit 
for XENON1T~\cite{Aprile:2015uzo}. Almost the entire parameter space ($m_\chi, m_V$) 
for the vector mediator case considered here is already ruled out from the LUX data. We find that for any DM 
mass the scattering cross-section generally increases as the degenerate parameter region is approached. 
This happens because of a resonant enhancement of $\sigma_{\rm SI}$ near r = 1.

\subsection{Indirect detection}
\label{sec:indirect}
The Fermi Large Area Telescope (LAT) collaborations~\cite{Ackermann:2015zua} have dedicated detectors to 
measure cosmic ray fluxes arising from DM annihilation in the Universe. 
In Figure~\ref{Figure:3} we show the prediction for the total DM annihilation into $u \bar{u}$ for the vector/scalar 
mediated $t$-channel model. The predictions shown here are for the DM mass, mediator mass and the couplings consistent with
the observed relic density. We have also shown the bounds from the 95\% CL upper limits on the thermally-averaged cross-section for 
DM particles annihilating into $u\bar u$ Fermi-LAT observations. As expected in the parameter region, where co-annihilation is
important, the $\chi\,\bar{\chi}$ annihilation cross-section in the $u\,\bar{u}$ channel is greatly suppressed.
Even in the region away from resonance the Fermi-LAT data does not provide strong bounds on the mass and coupling
parameters in the entire range consistent with $\Omega_{\rm DM}h^{2}\,=\,0.12$.
\section{Collider bounds}
\label{sec:collider}
The $t$-channel mediator model considered here has scalar and vector mediators which carry colour, $SU(2)_{L}$
and $U(1)$ charges. They can thus be singly produced in association with DM particles, or pair produced if they are 
light enough at the LHC. These processes will contribute to the monojet and dijet signals with missing energy, 
with distinct signatures that can be searched for in dedicated searches.     
For the monojet events $q\,g\,\rightarrow\,q\,\chi\,\bar{\chi}$ are the dominant processes, in comparison to 
$q\,\bar{q}\,\rightarrow\,g\,\chi\,\bar{\chi}$, because of the large parton distribution probability of the gluon, 
as compared to quark and antiquark in the proton. The authors of the simplified DM model document~\cite{Abdallah:2015ter} 
have emphasised that the dominance of the associated production channels is a distinct feature of $t$-channel models. 
The 8\,TeV CMS collaboration data based on an integrated luminosity 19.7\,fb$^{-1}$~\cite{Askew:2014kqa, Khachatryan:2014rra} 
has been used by the authors of Ref.~\cite{Goyal:2016zeh, An:2013xka} to put bounds on the coupling of fermionic DM 
as a function of the mediator and DM mass for the case of scalar and vector mediators. In the present study we confine 
ourselves to constraints arising from the monojet signals using the parameters space $(m_{\chi}\,m_{S/V} )$ for different 
values of the couplings $\left(g_{\chi}^{\rm S}\right)^{i}$/ $\left(g_{\chi}^{\rm V}\right)^{i}$/ $\left(c_{\chi}^{\rm V}\right)^{i}$
consistent with the observed relic density. The cross-section for 
monojet events is obtained by generating parton level events for the process $p\,p\,\rightarrow\,\chi\,\bar{\chi}\,j$ using 
\texttt{MadGraph}~\cite{Alwall:2014hca}, where the model file for the Lagrangian is obtained from \texttt{FeynRules}, and 
we use \texttt{CTEQ611} parton distribution function for five flavour quarks in the initial state. We employ the usual 
cuts, and the cross-sections are calculated to put bounds on the parameters of the model by requiring 
(i)  $E^{miss}_{T}>\,250\,$GeV and (ii) $E^{miss}_{T}>\,450\,$GeV, for which the CMS result excludes new contributions 
to the monojet cross-section for the scalar and vector mediators as shown in Figure~\ref{Figure:4} as function of $m_{\chi}$.
For the values of mediator mass $m_{S}/m_{V}$ consistent with the relic density. The results are displayed for some 
representative values of the couplings. From Figure~\ref{Figure:4} we find that the collider bounds are much weaker 
compared to the bounds from the direct detection experiments for the vector mediator case. The scalar mediator case is 
interesting in this case as the collider bound rules out low mass DM particles.The bounds from the monojet $+$ missing 
energy cross section puts a lower limit on the DM particle mass, where the limit depends on the coupling, and increases with 
the coupling.          
\section{Summary and discussion}
\label{sec:summary}
In this paper we have considered a spin-3/2 DM particle interacting with the SM fermions through the 
exchange of a scalar  or a vector mediator in the $t$-channel. Invoking MFV we restricted ourselves to the 
coupling of DM candidates with SM singlet right-handed quarks with universal coupling. The thermal relic DM abundance has been 
computed by taking into account the co-annihilation processes. Co-annihilation has the effect of reducing the Yukawa couplings needed
to generate the required DM density. The co-annihilation effects are more pronounced in the large $m_{\chi}$ regime, where 
mediator self annihilation into gauge bosons has the potential to suppress the relic density below the observed value.
Similar behaviour was observed in $t$-channel model for spin-1/2~\cite{Bai:2013iqa} and scalar
DM~\cite{Giacchino:2015hvk} particles. Our main observations are the following:
\begin{itemize}
\item[(a)] The direct detection experiments, through DM-nucleon elastic scattering data, provide
the most stringent bounds for the case of a vector mediator. In this case the entire parameter space 
allowed by the relic density is already ruled out by the LUX data. This result is consistent with the earlier studies of spin-3/2 DM
in the EFT~\cite{Yu:2011by} frame work for pure vector couplings, as well as in a simplified $s$-channel model~\cite{Khojali:2016pvu}.
The co-annihilation is unable to ameliorate this situation.

\item[(b)] There are no strong bounds from the the direct detection experiments on the scalar mediated interactions due to the 
velocity suppression of $\sigma^{\rm SI}$. In contrast, in the EFT frame work, both the scalar as well as vector interactions give rise
to dominant spin-independent nucleon-DM scattering cross-section and direct detection rules out scalar interaction for spin-3/2 DM
particles of mass lying between 10 GeV and 1 TeV~\cite{Yu:2011by}.

\item[(c)] The current constraints from indirect searches like, Fermi-LAT data, are not sensitive enough to put any meaningful 
constraints.

\item[(d)] Monojet searches at the LHC do not provide strong bounds at the vector couplings in comparison
to the bounds from direct detection experiments. However, for the case of the scalar mediator, where we do
not get any strong bounds from the direct detection experiment, collider bounds put a lower limit on the
DM mass which is $m_\chi \geq 300$~GeV. This limit rises with the increase in coupling.
\end{itemize}
Finally, it may be mentioned that bounds from direct detection experiments can, however, be evaded by foregoing the universal 
coupling between DM mediators and quarks, and letting the DM particles interact with only one generation, say with the third 
generation quarks (top-philic DM). 

\begin{acknowledgements}
MOK and ASC are supported by the National Research Foundation of South Africa.
\end{acknowledgements}


\begin{thebibliography}{9}
\bibitem{Ade:2015xua} 
P.~A.~R.~Ade {\it et al.} [Planck Collaboration],
Astron.\ Astrophys.\  {\bf 594}, A13 (2016)
[arXiv:1502.01589 [astro-ph.CO]].
\bibitem{Askew:2014kqa} 
A.~Askew, S.~Chauhan, B.~Penning, W.~Shepherd and M.~Tripathi,
Int.\ J.\ Mod.\ Phys.\ A {\bf 29}, 1430041 (2014)
[arXiv:1406.5662 [hep-ph]].
\bibitem{Akerib:2015cja} 
D.~S.~Akerib {\it et al.} [LZ Collaboration],
arXiv:1509.02910 [physics.ins-det].
\bibitem{Aprile:2015uzo} 
E.~Aprile {\it et al.} [XENON Collaboration],
JCAP {\bf 1604}, no. 04, 027 (2016)
[arXiv:1512.07501 [physics.ins-det]].
\bibitem{Tan:2016zwf} 
A.~Tan {\it et al.} [PandaX-II Collaboration],
Phys.\ Rev.\ Lett.\  {\bf 117}, no. 12, 121303 (2016)
[arXiv:1607.07400 [hep-ex]].
\bibitem{Akerib:2016vxi} 
D.~S.~Akerib {\it et al.} [LUX Collaboration],
Phys.\ Rev.\ Lett.\  {\bf 118}, no. 2, 021303 (2017)
[arXiv:1608.07648 [astro-ph.CO]].
\bibitem{Yu:2011by} 
Z.~H.~Yu, J.~M.~Zheng, X.~J.~Bi, Z.~Li, D.~X.~Yao and H.~H.~Zhang,
Nucl.\ Phys.\ B {\bf 860}, 115 (2012)
[arXiv:1112.6052 [hep-ph]].
\bibitem{Ding:2012sm} 
  R.~Ding and Y.~Liao,
  JHEP {\bf 1204}, 054 (2012)
  [arXiv:1201.0506 [hep-ph]].
\bibitem{Ding:2013nvx} 
  R.~Ding, Y.~Liao, J.~Y.~Liu and K.~Wang,
  JCAP {\bf 1305}, 028 (2013)
  [arXiv:1302.4034 [hep-ph]].
\bibitem{Savvidy:2012qa} 
  K.~G.~Savvidy and J.~D.~Vergados,
  Phys.\ Rev.\ D {\bf 87}, no. 7, 075013 (2013)
  [arXiv:1211.3214 [hep-ph]].
\bibitem{Dutta:2015ega} 
S.~Dutta, A.~Goyal and S.~Kumar,
JCAP {\bf 1602}, no. 02, 016 (2016)
[arXiv:1509.02105 [hep-ph]].
\bibitem{Chang:2017dvm} 
C.~F.~Chang, X.~G.~He and J.~Tandean,
arXiv:1704.01904 [hep-ph].
\bibitem{Khojali:2016pvu} 
M.~O.~Khojali, A.~Goyal, M.~Kumar and A.~S.~Cornell,
Eur.\ Phys.\ J.\ C {\bf 77}, no. 1, 25 (2017)
[arXiv:1608.08958 [hep-ph]].
\bibitem{Chang:2013oia} 
S.~Chang, R.~Edezhath, J.~Hutchinson and M.~Luty,
Phys.\ Rev.\ D {\bf 89}, no. 1, 015011 (2014)
[arXiv:1307.8120 [hep-ph]].
\bibitem{DiFranzo:2013vra} 
A.~DiFranzo, K.~I.~Nagao, A.~Rajaraman and T.~M.~P.~Tait,
JHEP {\bf 1311}, 014 (2013)
Erratum: [JHEP {\bf 1401}, 162 (2014)]
[arXiv:1308.2679 [hep-ph]].
\bibitem{An:2013xka} 
H.~An, L.~T.~Wang and H.~Zhang,
Phys.\ Rev.\ D {\bf 89}, no. 11, 115014 (2014)
[arXiv:1308.0592 [hep-ph]].
\bibitem{Papucci:2014iwa} 
M.~Papucci, A.~Vichi and K.~M.~Zurek,
JHEP {\bf 1411}, 024 (2014)
[arXiv:1402.2285 [hep-ph]].
\bibitem{Christensen:2013aua} 
N.~D.~Christensen {\it et al.},
Eur.\ Phys.\ J.\ C {\bf 73}, no. 10, 2580 (2013)
[arXiv:1308.1668 [hep-ph]].
\bibitem{Hassanain:2009at} 
B.~Hassanain, J.~March-Russell and J.~G.~Rosa,
JHEP {\bf 0907}, 077 (2009)
[arXiv:0904.4108 [hep-ph]].
\bibitem{E. W. Kolb}
E.~W.~Kolb nd M.~S.~Turner,
Front.~Phys.~{\bf 69}, 1(1990).
\bibitem{Belanger:2014vza} 
G.~Bélanger, F.~Boudjema, A.~Pukhov and A.~Semenov,
Comput.\ Phys.\ Commun.\  {\bf 192}, 322 (2015)
[arXiv:1407.6129 [hep-ph]].
\bibitem{Alloul:2013bka} 
A.~Alloul, N.~D.~Christensen, C.~Degrande, C.~Duhr and B.~Fuks,
Comput.\ Phys.\ Commun.\  {\bf 185}, 2250 (2014)
[arXiv:1310.1921 [hep-ph]].
\bibitem{DelNobile:2013sia} 
M.~Cirelli, E.~Del Nobile and P.~Panci,
JCAP {\bf 1310}, 019 (2013)
[arXiv:1307.5955 [hep-ph]].
\bibitem{Freytsis:2010ne} 
  M.~Freytsis and Z.~Ligeti,
  Phys.\ Rev.\ D {\bf 83}, 115009 (2011)
  [arXiv:1012.5317 [hep-ph]].
\bibitem{Barger:2008qd} 
V.~Barger, W.~Y.~Keung and G.~Shaughnessy,
Phys.\ Rev.\ D {\bf 78}, 056007 (2008)
[arXiv:0806.1962 [hep-ph]].
\bibitem{Ackermann:2015zua} 
M.~Ackermann {\it et al.} [Fermi-LAT Collaboration],
Phys.\ Rev.\ Lett.\  {\bf 115}, no. 23, 231301 (2015)
[arXiv:1503.02641 [astro-ph.HE]].
\bibitem{Abdallah:2015ter}
J.~Abdallah {\it et al.},
Phys.\ Dark Univ.\  {\bf 9-10} (2015) 8
[arXiv:1506.03116 [hep-ph]].
\bibitem{Khachatryan:2014rra} 
V.~Khachatryan {\it et al.} [CMS Collaboration],
Eur.\ Phys.\ J.\ C {\bf 75}, no. 5, 235 (2015)
[arXiv:1408.3583 [hep-ex]].
\bibitem{Goyal:2016zeh} 
A.~Goyal and M.~Kumar,
JCAP {\bf 1611}, no. 11, 001 (2016)
[arXiv:1609.03364 [hep-ph]].
\bibitem{Alwall:2014hca} 
J.~Alwall {\it et al.},
JHEP {\bf 1407}, 079 (2014)
[arXiv:1405.0301 [hep-ph]].
\bibitem{Bai:2013iqa} 
  Y.~Bai and J.~Berger,
  JHEP {\bf 1311}, 171 (2013)
  [arXiv:1308.0612 [hep-ph]].

\bibitem{Giacchino:2015hvk} 
  F.~Giacchino, A.~Ibarra, L.~Lopez Honorez, M.~H.~G.~Tytgat and S.~Wild,
  JCAP {\bf 1602}, no. 02, 002 (2016)
  [arXiv:1511.04452 [hep-ph]].
\end{thebibliography}
\end{document}